\documentclass{PoS}

\title{Real-time atmospheric monitoring for the Cherenkov Telescope Array using a wide-field optical telescope}

\ShortTitle{Real-time atmospheric monitoring for CTA using a wide-field optical telescope}

\author{\speaker{Jan Ebr}, Petr Jane\v{c}ek, Michael Prouza and Ji\v{r}\'{i} Bla\v{z}ek for the CTA Consortium\thanks{Full consortium author list at http://cta-observatory.org}\\
Institute of Physics of the Czech Academy of Sciences, Prague\\
E-mail: \email{ebr@fzu.cz}}

\abstract{The Cherenkov Telescope Array (CTA) is the next generation of ground-based very high energy gamma-ray instruments and is planned to be built on two sites (one in each hemisphere) in the coming years, with full array operation foreseen to begin 2020. The goal of performing high precision gamma-ray energy measurements while maximizing the use of observation time demands detailed and fast information about atmospheric conditions. Besides LIDARs designed to monitor clouds and aerosol content of the atmosphere in the pointing direction of the CTA telescopes, we propose to use the "FRAM" (F(/Ph)otometric Robotic Atmospheric Monitor) device, which is a small robotic astronomical telescope with a large field of view and a sensitive CCD camera that together ensure precise atmospheric characterization over the complete field-of-view of the CTA.
FRAM will use stellar photometry to measure atmospheric extinction across the field of view of the CTA without interfering with the observation (unlike laser-based methods). This allows FRAM to operate with high temporal resolution and provide both real-time data for on-the-fly scheduling decisions and an offline database for calibration and selection of scientific data. The fast robotic mount of the telescope supports quick observation of multiple fields when the array is split and even a check of the conditions in the directions of the upcoming observations is possible. The FRAM concept is built upon experience gained with a similar device operated at the Pierre Auger Observatory. A working prototype of FRAM proposed for CTA is being built in Prague for extensive testing before deployment on site; first results and experiences with this prototype are presented.
}

\FullConference{The 34th International Cosmic Ray Conference,\\
30 July- 6 August, 2015\\
The Hague, The Netherlands}

\begin{document}

\section{Introduction}

The Cherenkov Telescope Array (CTA) is expected to enter full operation in 2020 as the next generation of ground-based very high energy gamma-ray instruments \cite{CTA}. Its unprecedented scientific potential comes not only from the sheer size, with tens of telescopes of different sizes across two sites (one in each hemisphere), but also from the strict requirements established on the precision of measurements and efficiency of use of the observation time. To meet these requirements, careful calibration of the instrument is key, including precise knowledge of momentary atmospheric conditions, in particular the presence of clouds and the aerosol content of the atmosphere. From the devices proposed to be installed at the CTA sites \cite{MG}, the LIDARs can measure clouds and aerosols with high accuracy in a given direction \cite{Lidar} and the All-Sky Cameras can monitor the whole sky at all times without interfering with the CTA observations \cite{ASC} -- however neither of those devices can provide detailed information about the conditions in the current CTA field of view with high spatial and temporal resolution. Thus we propose a complementary device -- a small robotic optical telescope, the "FRAM" (F(/Ph)otometric Robotic Atmospheric Monitor), which will measure the transparency of the atmosphere in the direction of the CTA field of view using stellar photometry. In isolation, this method is not without shortcomings, because only the integral extinction from ground to the upper edge of the atmosphere can be provided (without a detailed altitude profile), but if operated together with profiling atmospheric devices (like to LIDARs), it will help provide a complete picture of the atmospheric conditions across the CTA field of view, in real time during observations.

The CTA FRAM concept draws upon experience gained over a decade with a pioneering device \cite{FRAM} that has been operated at the Pierre Auger Observatory \cite{Auger} -- more details on the current implementation of the Auger FRAM can be found in \cite{Malaga2} and an explanation of its use for rapid atmospheric monitoring at Auger in \cite{Malaga1}. The differences between the application of stellar photometry at Auger and CTA as well as the changes in the hardware setup that follow from them are explained in detail in \cite{CTAFRAM}. In this paper, we briefly review the current status of the design of the FRAM for CTA and then we discuss the performance of the partial prototype, which has been assembled at the Institute of Physics in Prague and has received first light at the beginning of June 2015, and advances in data analysis. 

\section{Hardware and software}

We currently foresee having three identical FRAM devices at the CTA sites (one in the northern site, two in the larger southern site to cover for spatial variations in atmospheric conditions over the extended area). Based on the extensive testing carried out over years at the Auger FRAM, we have chosen components that have proven reliable and efficient. Apart from the housing and roof, which are designed and built according to our specifications by local specialist companies, and some custom structural elements, all the other components are readily available commercial products -- a fact that has allowed us to have a working partial prototype (Fig.~\ref{fig-1}) within a couple of months from receiving funding for it. The main components of the prototype (and thus of the proposed future CTA FRAMs) are:

\begin{figure}[h!t]
\centering
\includegraphics[width=12cm,clip]{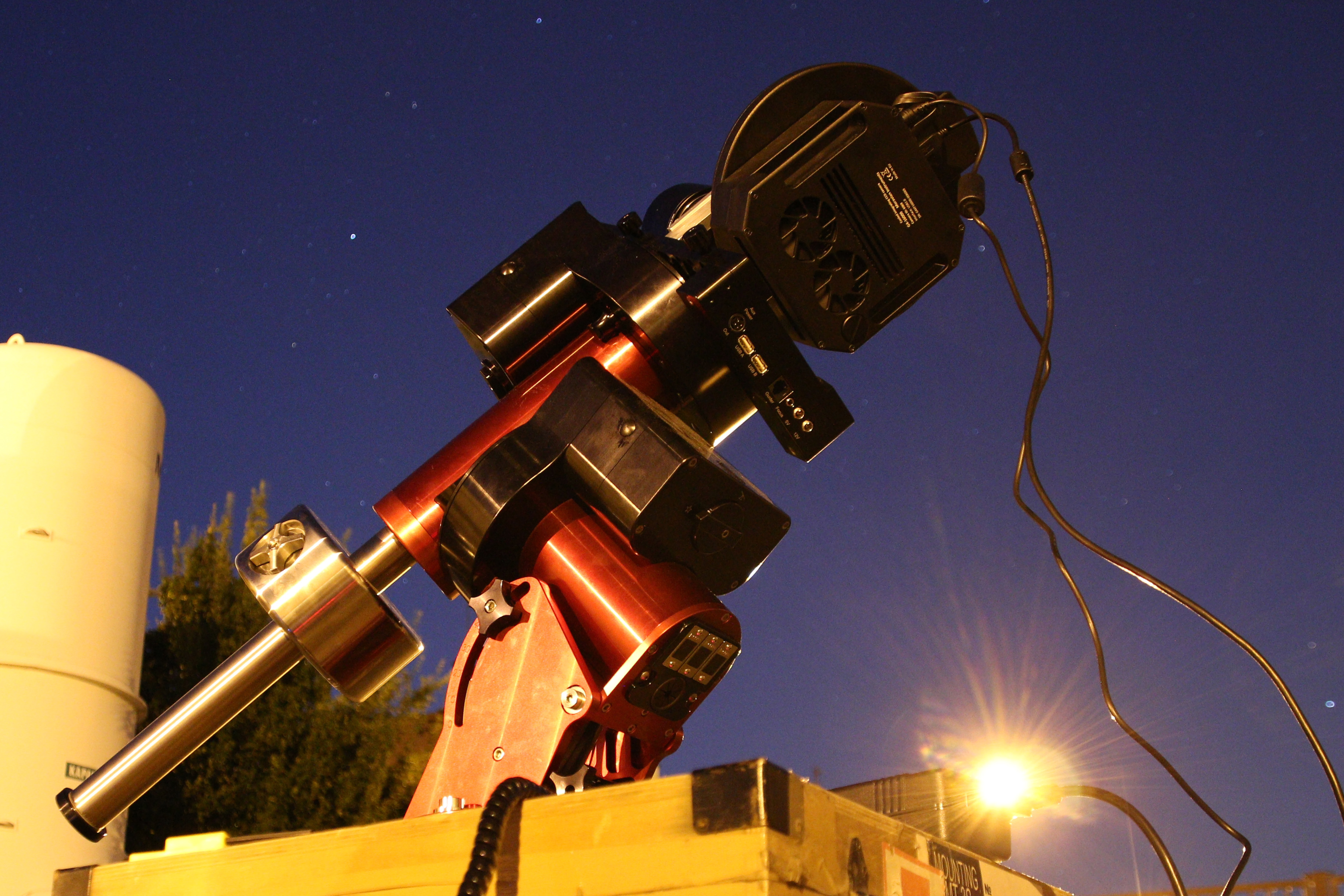}
\caption{The partial prototype of the FRAM for CTA (without housing, pillar and dedicated computer) receives its first light in the backyard of the Institute of Physics in Prague.}
\label{fig-1}
\end{figure}

\begin{itemize}
\item{Paramount MYT robotic astronomical mount. The combination of camera and lens used is lightweight and could be easily used with a less robust equatorial mount; however the Paramount mounts have the clear advantage of time-proven reliability and ability to autonomously regain orientation after a power cut or loss of communication thanks to their absolute homing system. }
\item{Moravian Instruments G4-16000 large-format CCD camera equipped with standard BVRI photometric filters. The KAF-16803 sensor used in this camera spans 36$\times$36-mm, and thus allows the FRAM to reach large field of view while keeping a relatively big physical aperture of the lens. Collecting enough starlight in a short time window is crucial for the envisaged high temporal resolution of the measurements.}
\item{Zeiss 135 f/2.0 photographic lens. This particular combination of sensor size and lens focal length provides a field of view of $15\times15$ degrees and thus covers the whole expected CTA field of view (expected to be up to 10 degrees) with a sufficient margin for the parallax with respect to the more distant telescopes. The Zeiss lens is reportedly unrivaled in sharpness at full aperture at this focal length, particularly considering the large sensor area to cover -- the first light results have even exceeded our expectations in this respect.}
\item {An external focuser which mechanically couples to the focus ring of the lens and thus allows remote re-focusing; this component has not yet been delivered, but non-remote operation of the prototype is easily done using manual focus.}
\end{itemize}

\begin{figure}[h!t]
\centering
\includegraphics[width=11cm,clip]{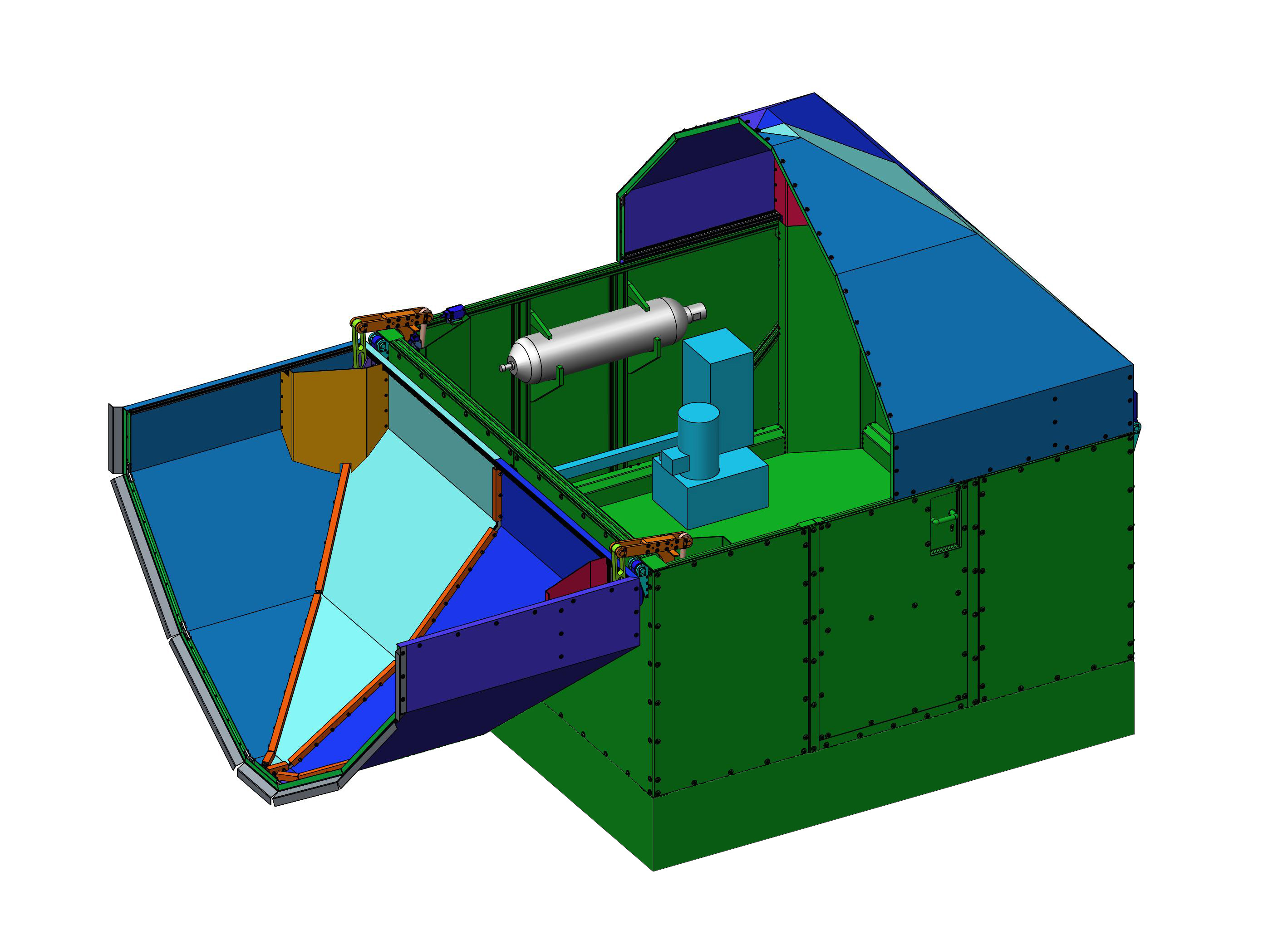}
\caption{3D model of the house and roof}
\label{fig-2}
\end{figure}

Eventually, each FRAM will be located in a small house with a roof that fully unfolds in the night to allow unobstructed observation of the full sky. The design of the housing (by Elya Solutions) is now finished (Fig.~\ref{fig-2}) and the production of the first house and roof has started. We have chosen a footprint that is rather large with respect to the FRAM device ($2.5\times2.5$ meters) to allow reasonable maintenance access to FRAM with the dome closed for environmental protection. With such a large footprint already set, the design of the roof was chosen to be flexible so that it can be used for other robotic telescopes -- the housing is suitable for any telescope that fits into a sphere with a radius of 80 cm during operation (for example a Schmidt-Cassegrain with 30 cm diameter as in the Auger FRAM).

Both the walls and the moveable roof will be made of aluminium composite panels (a sandwich structure of plastic between two aluminium sheets) reinforced with aluminium profiles; all parts can be easily bolted together and to the concrete foundations on site. The roof will be opened and closed using a hydraulic system with electric valves controlled by the Schneider Zelio controller. In case of a power outage, a pressure storage reservoir (hydraulic accumulator) will provide the required pressure to close the dome, while the valves will run from a battery. For additional safety, a manual hydraulic pump and mechanical valve controls will be also provided. When closed, the roof will be weather-proof and its operation is possible with winds up to 60 km/h; it can also be opened even after heavy snowfall without danger of snow falling on the telescope.

While the computationally demanding image processing would need to be handled by a server in the CTA data centre, a small local computer is necessary to control the mount, camera, focuser, roof and auxiliary devices. Intel DE3816TYKHE, a low-power, passively cooled industrial computer was chosen because of its relatively strong-for-its-size I/O capabilities (6 USB ports, gigabit Ethernet), necessary for controlling all the connected devices and for quick transfer of the large images from CCD camera (32 MB) to the data centre. It will run Linux and the RTS2 software \cite{RTS} for robotic observatories.  The communication of FRAM with the central CTA array control system would be implemented (similarly to many other CTA devices) using the OPC UA interface.

\section{First light and image processing}

\begin{figure}
\centering
\includegraphics[width=13cm,clip]{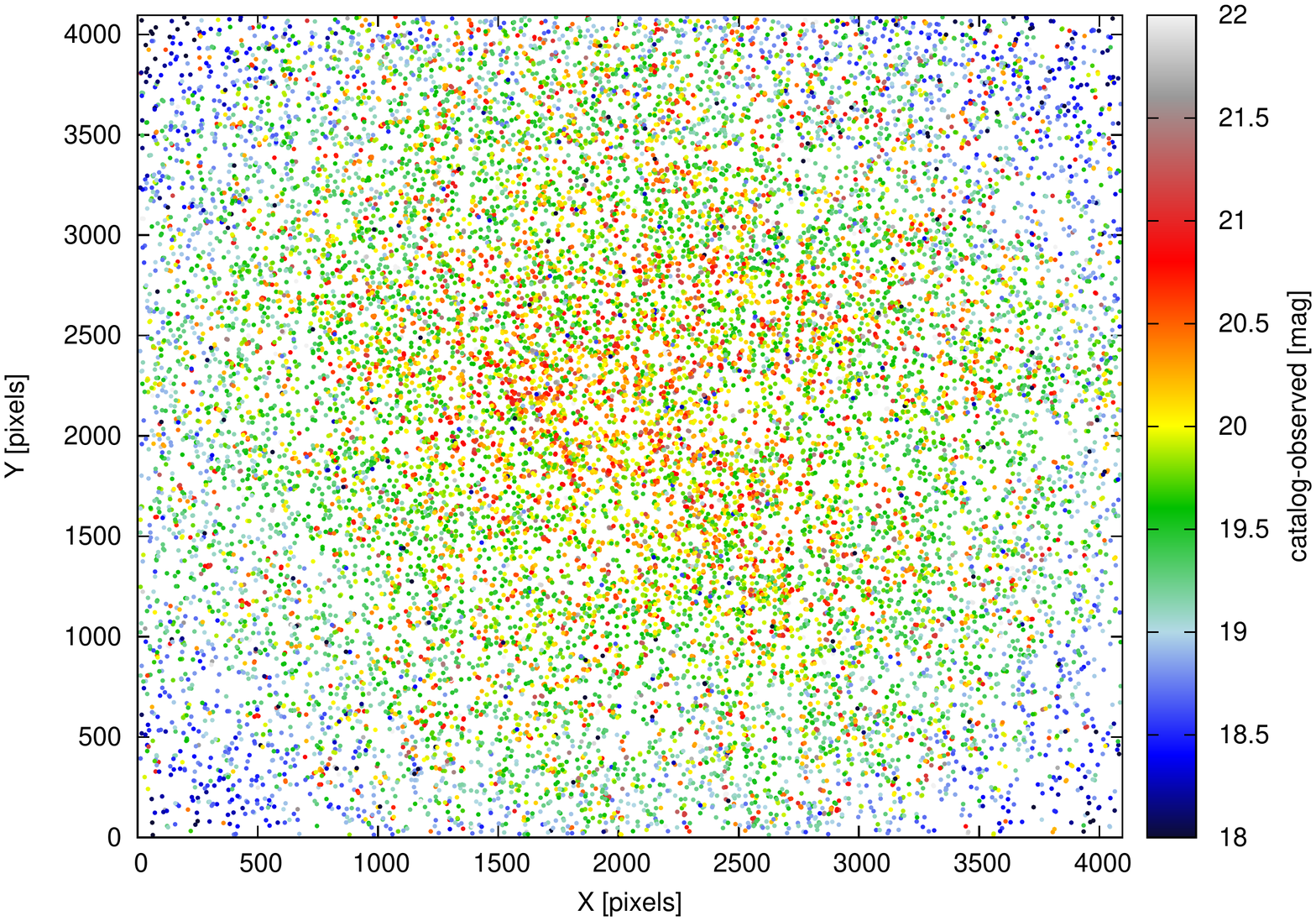}
\caption{All identified stars from an image taken within the milky way in Cygnus. The color code shows the difference between measured and catalog brightness for each star; the value is plotted before zero-point calibration and thus has an arbitrary offset. Note the vignetting clearly visible in the corners and the residual artifacts from the 6$\times$6 matrix of sub-images used for star identification (stars cannot be identified very close to the border of a sub-image).}
\label{fl1}
\end{figure}

During a very clear night at the beginning of June 2015 we have taken first images of the sky with the FRAM prototype. Here we use them to demonstrate the methods of image processing. All images shown are taken in the V filter with a 15-second exposure.

Because image distortion over the large field of view makes star identification difficult, each image taken is first split into a matrix of 6$\times$6 sub-images. Across a single sub-image, the distortion is negligible and thus the relation between the position of a star on the image in pixels and its celestial coordinates can be taken as linear. For each point source detected in the image, the apparent brightness is measured and the Tycho catalog \cite{Tycho} is searched for a corresponding star (with preference to brighter stars when more catalog entries are very close to each other). For a random field within the milky way in Cygnus (Cyg), about 20 thousand stars were identified; when observing perpendicularly to the galactic plane in Ursa Major (UMa), we detected roughly 4000 stars.

The lens produces useful stellar images at full aperture even in the corners of the CCD chip (which is slightly larger than 35 mm film), however there is significant vignetting (up to 2 magnitudes) in the corners (Fig.~\ref{fl1}). While the vignetting can be removed using flat-fielding, it greatly reduces precision of the measurement in the affected areas; however these are only a very small part of the field of view. 

\begin{figure}
\centering
\includegraphics[width=13.5cm,clip]{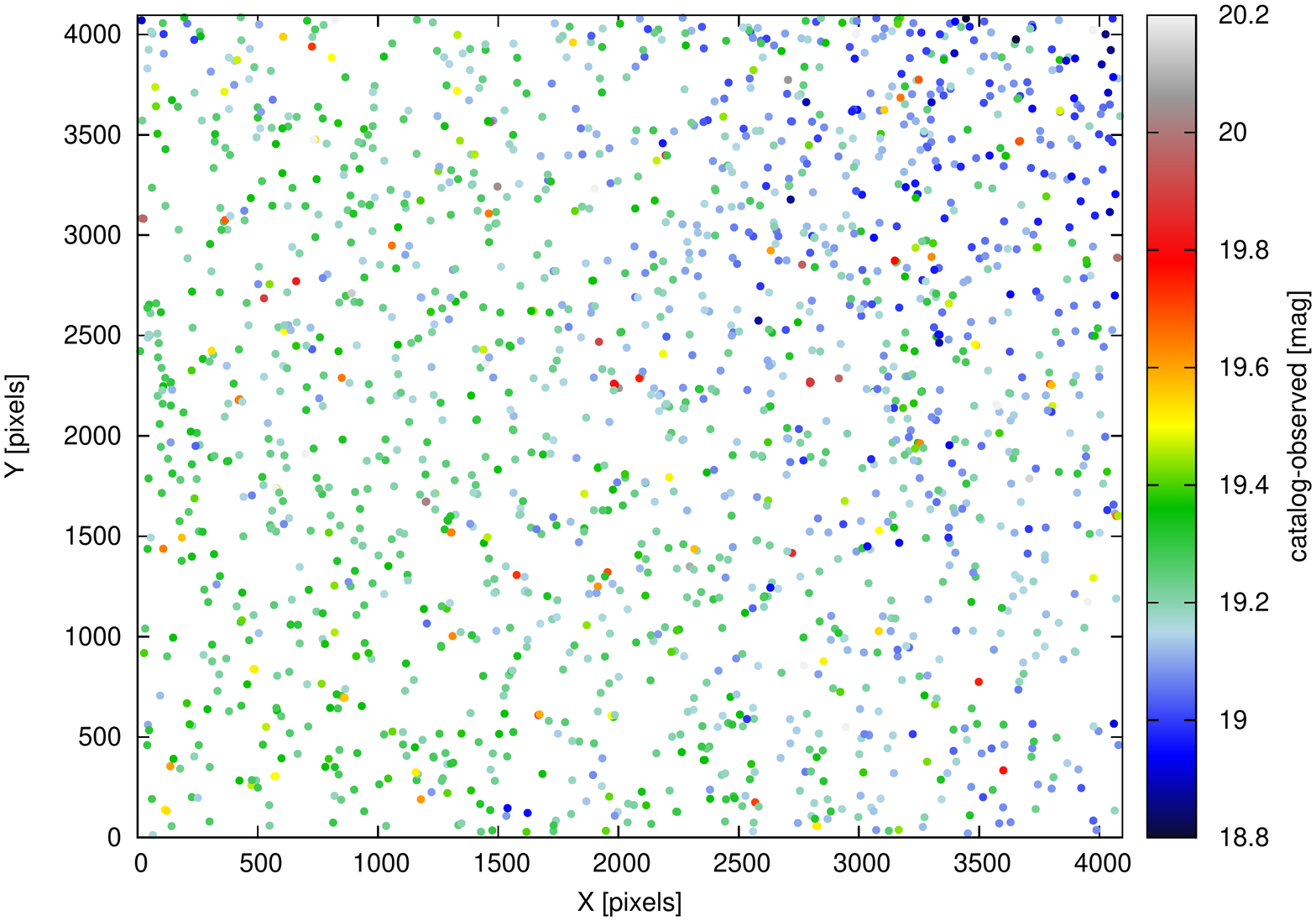}
\caption{Identified stars brighter than 9.5 mag from a flat-field corrected image. The visible gradient is due to the zenith-angle dependence of the extinction (the image spans altitudes between roughly 15 to 30 degrees above the horizon). }
\label{fl2}
\end{figure}

Fig.~\ref{fl2} shows the same image after applying the flat-field correction and selecting only stars brighter than 9.5 mag, for which both the catalog and measured brightness are reasonably precise (1825 such stars in the Cyg field and 638 stars in the UMa field). The spread of the values of the difference between catalog and measured brightness is visibly smaller (note the different scale), but a gradient remains, due to the dependence of the atmospheric extinction on the traversed airmass $X$ (and thus the zenith angle), which is for such a large field already clear from one image. If we express catalog vs. observed brightness as 
$$m_{obs}=m_{cat}+a+kX$$
and we fit the zero-point $a$ and the extinction coefficient $k$ within the image, we obtain a reasonable value of $k=0.075$ for the night of observation.

\begin{figure}
\centering
\includegraphics[width=13.5cm,clip]{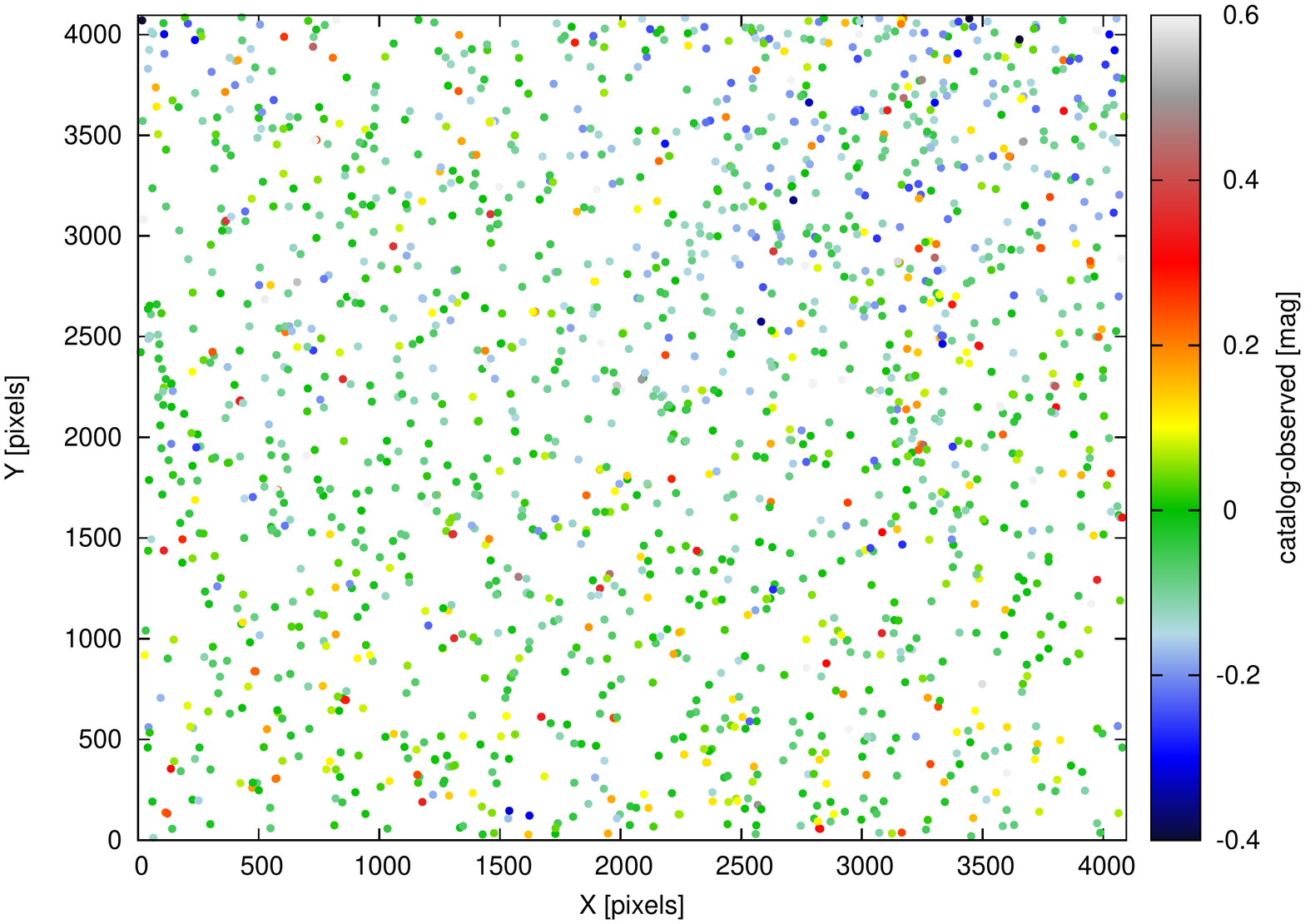}
\caption{Identified stars brighter than 9.5 mag from an image corrected for flat-field and the dependence of atmospheric extinction on airmass. The extinction fit also removes the arbitrary offset due to the zero-point of the camera.}
\label{fl3}
\end{figure}

Using this fit, we can correct for atmospheric extinction (and apply the obtained zero-point to convert the magnitude scale), as shown in Fig.~\ref{fl3}. Here we see a residual gradient (likely due to inadequate flat-field caused by parasitic light from nearby artificial sources) and some outliers (likely due to mis-identified stars, catalog errors, and stars with large color indices), but for most of the stars, the difference between catalog and observed brightness lies within 0.1 mag from the mean value; we expect further improvements after we perform detailed corrections for star colors and the spectral response of the system, as well as more careful flat-fielding of the camera.

\begin{figure}
\centering
\includegraphics[width=13.5cm,clip]{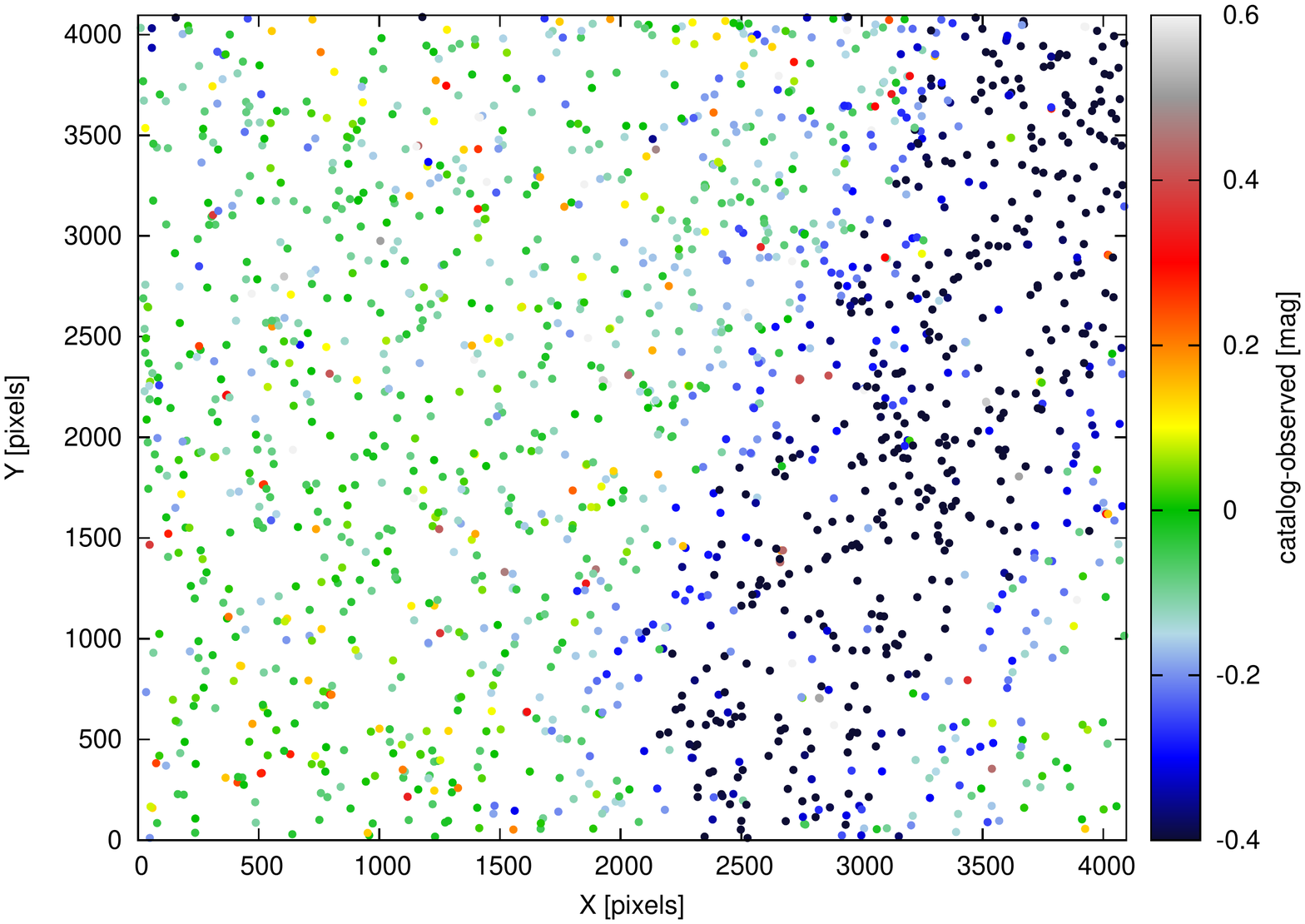}
\caption{An example of an "artificial thin cloud" detected using stellar photometry.}
\label{fl4}
\end{figure}

Ironically, the very clear weather during the time available for testing prevented us from detecting actual clouds with the prototype. As a substitute, we have created an "artificial thin cloud" by hanging a dark cloth across the field of view for 5 seconds during a 15-second exposure. This method simulated a cloud or other source of absorption which does not remove the stars completely and thus cannot be detected by simply counting the number of identified sources vs. the expected number from the catalog. Fig.~\ref{fl4} shows such an "artificial cloud" as a decrease of observed brightness of a large number of stars.

\section{Conclusions}

We have introduced the FRAM device proposed for atmospheric monitoring at the future CTA sites using stellar photometry. We have shown using the partial prototype of the FRAM that the technique can deliver detailed information on atmospheric extinction across a large field of view without interfering with the operation of other optical detection devices. This information can be used by the CTA to evaluate the atmospheric conditions, make operative decisions based on the current situation and to assess the quality of data and select those that are not significantly affected by atmospheric irregularities. 

\section*{Acknowledgements}

We gratefully acknowledge support from the agencies and organizations listed under Funding Agencies at this website: http://www.cta-observatory.org/. The authors are grateful for the support by the grants of the Ministry of Education of the Czech Republic MSMT-CR LE13012 and LG14019.

\end{document}